\documentclass[aps,prl,twocolumn,superscriptaddress,showpacs]{revtex4}
\usepackage{amsfonts}
\usepackage{amssymb}
\usepackage{amsmath}
\usepackage{graphicx}

\usepackage{dcolumn}   
\usepackage{bm}        
\usepackage{flafter}
\usepackage[dvips]{color}

\begin{document}

\title{Optical Study of the Free Carrier Response of LaTiO$_{3}$/SrTiO$_{3}$ Superlattices}
\date{\today}

\author{S. S. A. Seo}
\affiliation{ReCOE$\&$FPRD, Department of Physics and Astronomy,
Seoul National University, Seoul 151-747, Korea}
\author{W. S. Choi}
\affiliation{ReCOE$\&$FPRD, Department of Physics and Astronomy,
Seoul National University, Seoul 151-747, Korea}
\author{H. N. Lee}
\affiliation{Materials Science and Technology Division, Oak Ridge
National Laboratory, Oak Ridge, Tennessee 37831, USA}
\author{L. Yu}
\affiliation{Department of Physics and Fribourg Center for
Nanomaterials, University of Fribourg, 1700 Fribourg, Switzerland}
\author{K. W. Kim}
\affiliation{Department of Physics and Fribourg Center for
Nanomaterials, University of Fribourg, 1700 Fribourg, Switzerland}
\author{C. Bernhard}
\affiliation{Department of Physics and Fribourg Center for
Nanomaterials, University of Fribourg, 1700 Fribourg, Switzerland}
\author{T. W. Noh}\email{twnoh@snu.ac.kr}
\affiliation{ReCOE$\&$FPRD, Department of Physics and Astronomy,
Seoul National University, Seoul 151-747, Korea}

\begin{abstract}
We used infrared spectroscopic ellipsometry to investigate the
electronic properties of LaTiO$_{3}$/SrTiO$_{3}$ superlattices
(SLs). Our results indicated that, independent of the SL
periodicity and individual layer-thickness, the SLs exhibited a
Drude metallic response with sheet carrier density per interface
$\approx3\times10^{14}$ cm$^{-2}$. This is probably due to the
leakage of $d$-electrons at interfaces from the Mott insulator
LaTiO$_{3}$ to the band insulator SrTiO$_{3}$. We observed a
carrier relaxation time $\approx35\ $fs and mobility $\approx35\
$cm$^{2}$V$^{-1}$s$^{-1}$ at 10 K, and an unusual temperature
dependence of carrier density that was attributed to the
dielectric screening of quantum paraelectric SrTiO$_{3}$.

\end{abstract}
\smallskip


\maketitle

Recent advances in the growth of atomic-scale multilayers of
perovskites have opened up new avenues for tailoring their
electromagnetic properties. For example, Ohtomo \textit{et al.}
grew superlattices (SLs) consisting of an LaTiO$_3$ (LTO) Mott
insulator and an SrTiO$_3$ (STO) band insulator with atomically
abrupt interfaces (IFs) \cite{Ohtomo}. They observed an
interesting charge modulation involving electron transfer across
the IF from the LTO to the STO layers. This had a decay length of
1.0$\pm$0.2 nm, which is about one order of magnitude larger than
expected for conventional Thomas-Fermi screening. Subsequent
transport measurements have been interpreted in terms of an
unusual metallic state with quasi two-dimensional properties. This
intriguing experimental observation has stimulated a number of
theoretical investigations \cite{Okamoto_nature, Okamoto1,
theories, Kancharla}, which confirm that so-called ``electronic
reconstruction" \cite{Okamoto_nature} can indeed give rise to a
unique interfacial metallic state.

\begin{figure}
\includegraphics[width=3.3in,bb=20 20 407 218]{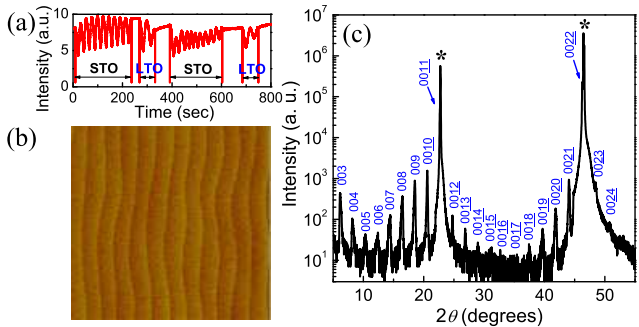}
\caption{\label{fig1}(color online) (a) RHEED intensity
oscillation during the initial growth of the (LTO)2(STO)10 SL. (b)
atomic force microscope topographic image (5$\times$5 $\mu$m$^2$)
of an (LTO)1(STO)10 SL 100 nm thick with single unit cell terraces
conserved. (c) X-ray \textit{$\theta$}-2\textit{$\theta$} scan of
(LTO)1(STO)10. Peaks marked with (*) in (c) are reflected from the
STO substrate.}
\end{figure}

There is now a great deal of demand for experimental confirmation
and direct investigation of such interfacial metallic states.
Takizawa \textit{et al.} recently measured photoemission spectra
of LTO/STO SLs with a topmost STO layer of variable thickness
\cite{Takizawa}. They observed a metallic Fermi edge, indicating
the formation of a metallic IF. However, to date there have been
no reports of quantitative experimental information on the
electrodynamic properties of the itinerant electrons. Here, we
present spectroscopic ellipsometry measurements of the infrared
(IR) dielectric properties for a set of (LTO)$\alpha$/(STO)10 SLs
with $\alpha$=1, 2, 4, and 5 unit cells of LTO layer and 10 unit
cells of STO. Our IR optical data yield reliable information about
the intrinsic electrodynamic properties of these SLs. First, they
clearly demonstrate that these SLs contain a sizeable
concentration of itinerant charge carriers. Furthermore, they
establish that the sheet carrier density is proportional to the
number of IFs, and its absolute value agrees closely with the
theoretical predictions. Our data also highlight an unusually
strong temperature ($T$) dependence of the carrier density, which
was unexpected for bulk metals, but can be explained in terms of
the $T$-dependent dielectric screening of the STO layer
controlling charge transfer across the LTO/STO IF.

We grew high-quality LTO/STO SLs $\sim$100 nm thick with
atomically flat surfaces and abrupt IFs on single crystalline STO
(001) substrates. To do this, we used pulsed laser deposition
(PLD) at $T$=720 $^\circ$C in P$_{O2}$=10$^{-5}$ Torr with
\textit{in situ} monitoring of the specular spot intensity of
reflection high-energy electron diffraction (RHEED) (see Ref.
\cite{HNLee} for more details on PLD growth). The RHEED intensity
oscillations in Fig. 1(a) confirmed the controlled growth of
alternating LTO and STO layers. Doing this at a higher oxygen
pressure would result in the growth of unwanted La$_2$Ti$_2$O$_7$
phases, as reported previously \cite{LTO227}. After growth, all
samples were exposed immediately to a higher pressure
(P$_{O2}$=10$^{-2}$ Torr) for \textit{in situ} postannealing at
growth \textit{T} for 5 min, and then cooled to room $T$. No
changes in the RHEED specular spot pattern were observed after
postannealing. The atomic force microscope topographic image in
Fig. 1(b) shows that the SLs retained their single unit cell
terrace steps even after deposition of $\sim$250 unit cells of LTO
and STO. Figure 1(c) shows x-ray
\textit{$\theta$}-2\textit{$\theta$} diffraction, exhibiting
well-defined satellite-peaks, confirming the high IF quality and
SL periodicity of LTO and STO layers. The full-width half-maximum
$\leq0.04^{\circ}$ in \textit{$\omega$}-scan, which is almost
identical to the substrate scan, confirmed the high crystallinity
of the SLs.

\begin{figure}
\includegraphics[width=3.3in,bb=20 20 404 222]{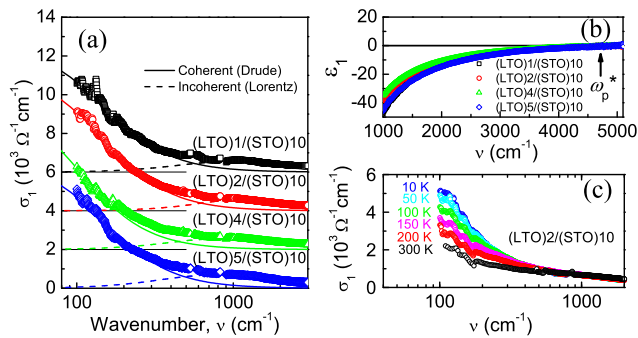}
\caption{\label{fig2}(color online) (a) Optical conductivity
spectra of (LTO)$\alpha$/(STO)10 at 10 K (with vertical offsets),
containing the coherent (Drude) and incoherent (Lorentz)
contributions. (b) SL real dielectric constant spectra as a
function of photon wavenumber at 10 K. The arrow indicates the
screened plasma frequency
$\omega_{p}^{*}[=\omega_{pD}/\sqrt{\varepsilon_{\infty}}]$. (c)
$T$-dependent $\sigma_1(\omega)$ of (LTO)2/(STO)10.}
\end{figure}

We measured the $T$-dependent IR optical properties of the SLs
using ellipsometry in the range of 80-5000 cm$^{-1}$ (10-600 meV).
The far-infrared (FIR) measurements for 80-700 cm$^{-1}$ were
performed with a home-built ellipsometer attached to a Bruker 66V
Fourier transform IR (FT-IR) spectrometer at the IR beamline of
the ANKA synchrotron at FZ Karlsruhe, Germany \cite{Bernhard}. For
the mid-infrared (MIR) range of 400-5000 cm$^{-1}$, we used a
home-built ellipsometer in combination with the glow-bar source of
a Bruker 113V FT-IR spectrometer. The FIR (MIR) measurement was
performed with an angle of incidence of the linearly polarized
light of 82.5$^{\circ}$ (80$^{\circ}$), $i.e.$ close to the
pseudo-Brewster angle of the SLs. Details about the ellipsometry
technique are given in Ref. \cite{Bernhard, Azzam}. Here, we only
point out that the ellipsometry is a self-normalizing technique
that directly measures the complex dielectric function
$\tilde{\varepsilon}$($\omega$)[=\emph{$\varepsilon$}$_1$(\emph{$\omega$})+\emph{$i\varepsilon$}$_2$(\emph{$\omega$})],
without the need for Kramers-Kronig analysis. We also measured the
$T$-dependent spectroscopic response of a bare STO substrate that
was treated thermally under the same growth and annealing
conditions of $T$ and P$_{O2}$ as the SLs. Then we used a
uniaxially anisotropic single-layer model \cite{Azzam} to obtain
the in-plane component \cite{c-axis} of
\emph{$\tilde{\varepsilon}$}(\emph{$\omega$}), and the related
optical conductivity
\emph{$\tilde{\sigma}$}(\emph{$\omega$})[=\emph{$\tilde{\varepsilon}$}(\emph{$\omega$})$\omega$/4$\pi$\emph{i}=\emph{${\sigma}$}$_1$(\emph{$\omega$})-\emph{i}\emph{$\varepsilon$}$_{1}$(\emph{$\omega$})$\omega$/4$\pi$]
of the SLs. As the thickness of the SLs is well below our IR
wavelength, the entire SL film can be treated as a single layer
according to the effective medium theory. The resulting effective
dielectric functions thus correspond to the volume-averaged
dielectric functions of the components in the SL. Our optical
technique can provide intrinsic values of the bulk properties of
these SLs, which can be compared with the preexisting results of
transport measurements \cite{Ohtomo}.

\begin{table}
\caption{\label{tab:table1}Drude-Lorentz fitting parameters for
optical spectra of (LTO)$\alpha$/(STO)10 SLs measured at 10 K.}
\begin{ruledtabular}
\begin{tabular}{cccccc}
 &\multicolumn{2}{c}{Drude}&\multicolumn{2}{c}{Lorentz}&\\
$\alpha$&$\omega^{2}_{pD}$ (cm$^{-1}$)&$\Gamma_{D}$ (cm$^{-1}$)&$S$&$\omega_{0}$ (cm$^{-1}$)&$\Gamma$ (cm$^{-1}$)\\
\hline
1&$6.0\times10^7$&$128(\pm3)$&$75(\pm11)$&$920(\pm95)$&$1980(\pm580)$\\
2&$6.6\times10^7$&$144(\pm8)$&$51(\pm22)$&$980(\pm240)$&$1530(\pm1200)$\\
4&$5.1\times10^7$&$105(\pm6)$&$87(\pm20)$&$780(\pm200)$&$1630(\pm1000)$\\
5&$6.1\times10^7$&$122(\pm2)$&$126(\pm8)$&$750(\pm65)$&$1740(\pm350)$\\
\end{tabular}
\end{ruledtabular}
\end{table}

Figures 2(a) and 2(b) show the low-$T$ spectra of
$\sigma_{1}(\omega)$ and $\varepsilon_{1}(\omega)$ for the series
of (LTO)$\alpha$/(STO)10 SLs with $\alpha$=1, 2, 4, and 5.
Significantly, all the spectra exhibited a prominent Drude-like
peak at low frequency in \emph{${\sigma}$}$_1$(\emph{$\omega$})
and a strongly inductive response in
\emph{${\varepsilon}$}$_1$(\emph{$\omega$}), which provided
unambiguous evidence that these SLs contained sizeable
concentrations of itinerant charge carriers. The strongly
conducting response also seemed to damp out IR active phonons in
the spectra. For quantitative analysis we applied the
Drude-Lorentz fitting function:
\begin{equation}
\tilde{\varepsilon}(\omega)=
\varepsilon_{\infty}-\frac{\omega^{2}_{pD}}{\omega^{2}+i\omega\Gamma_{D}}
+\frac{S\omega^{2}_{0}}{\omega^{2}_{0}-\omega^{2}-i\omega\Gamma}.
\label{eq:eq1}
\end{equation}
This is  shown in Fig. 2(a) by the solid and dashed lines. The
parameters of the Drude term are the scattering rate $\Gamma_{D}$,
and the plasma frequency $\omega^{2}_{pD}=4\pi{n}{e}^{2}/m^{*}$,
where \emph{n}, \emph{e}, and $m^{*}$ are the density, charge, and
effective mass of the itinerant charge carriers, respectively. The
parameters of the Lorentz term that accounts for the broad MIR
band are the oscillator strength $S$, the width $\Gamma$, and the
resonant frequency $\omega_{0}$. A broad MIR band is commonly
observed in conducting oxides and has been interpreted in terms of
either the inelastic interaction of the itinerant carriers, a
second type of carrier in a bound state, or low-lying interband
transitions. As LTO is well-known to have weak broad MIR bands,
the Lorentz peak with a small value of $S$ can be reasonably
interpreted to be the lowest intersite $d$-$d$ ($i.e.$, U-3J)
transition of LTO \cite{JSLee}. An overview of the fitting
parameters for the low-$T$ spectra is given in Table 1.

Most importantly, we noted that all the SLs have surprisingly high
$\omega_{pD}$, and thus high densities of free carriers. This
observation motivated us to follow up on suggestion of Okamoto and
Millis \cite{Okamoto_nature} of a quasi-two-dimensional metallic
state that develops in the LTO/STO IFs. We derived the sheet
carrier density per IF, $n_{sheet}=n\frac{d}{N_{I}}$, with the
total SL thickness $d$ and the number of LTO/STO IFs $N_{I}$ of
the SL \cite{LTO1}. Figure 3(a) shows $n_{sheet}$ values within
error-bars dependent on the detailed assignment of $m^{*}$ based
on the analysis of our ellipsometric data in the framework of the
extended Drude-model. This directly yields the renormalized mass
$m^{*}/m_{e}=(\omega_{p}/\omega)^{2}(\varepsilon_{0}-\varepsilon_{1})/((\varepsilon_{0}-\varepsilon_{1})^{2}-\varepsilon_{2}^{2})$.
Note that all of our SLs exhibit a very similar value of
$n_{sheet}\approx3\times10^{14}$ cm$^{-2}$ with $m^{*}=1.8m_e$
\cite{Okuda}, which agrees well with theoretical predictions
\cite{Okamoto1, Kancharla, integral}.

\begin{figure}
\includegraphics[width=3.3in,bb=20 20 409 191]{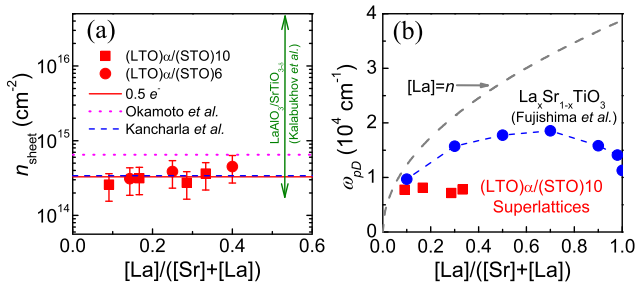}
\caption{\label{fig3}(color online) (a) $n_{sheet}$ per IF. The
solid line indicates the value of an ideal 0.5 $e^-$ per unit cell
area, and the dashed and dotted lines represent theoretically
predicted values reported by Kancharla \cite{Kancharla} and
Okamoto \cite{Okamoto1}, respectively. The vertical arrow shows
the range of $n_{sheet}$ in LaAlO$_3$/SrTiO$_{3-\delta}$ IF from
Ref. \cite{Kalakukhov}. (b) Comparison of SL $\omega_{pD}$ with
solid-solution \cite{Fujishima} and the ideal doping of
[La]=\emph{n} as a function of La and Sr composition.}
\end{figure}

Here, we should mention that the Drude term is only a
phenomenological description, and its microscopic origin is not
clear. One possible explanation is in terms of a chemical
intermixing of La and Sr ions across the IFs because
La$_x$Sr$_{1-x}$TiO$_3$ solid solutions are metallic
\cite{Tokura}. As indicated by the solid circles in Fig. 3(b), the
experimental data of bulk La$_x$Sr$_{1-x}$TiO$_3$ solid solutions
reported by Fujishima \textit{et al.} \cite{Fujishima} show that
the $\omega_{pD}$ value increases from x = 0.1 to 0.7, while it
decreases above 0.7 due to the strong electron correlation effect
as a function of increasing $x$. However, our SL data follow a
different path although the value of $\omega_{pD}$ is expected to
increase when going from $\alpha$ = 1 to $\alpha$ = 5, as shown by
the solid squares in Fig. 3(b). Moreover, the well-defined
superstructure peaks in our x-ray diffraction data and the abrupt
IFs seen on cross-sectional transmission electron microscopy (not
shown) confirm that the extent of La and Sr intermixing is
negligible.

As other candidates of extrinsic origin, it is also necessary to
consider the possibility of oxygen excess LaTiO$_{3+\delta}$
\cite{Schmehl} or La vacancy La$_{1-x}$TiO$_3$ \cite{Crandles2} of
LTO, and oxygen deficiency of STO \cite{Crandles1}. The former
possibility can be excluded as our data for $\omega_{pD}$ do not
exhibit a corresponding change as the relative volume fraction of
the LTO layers is altered by a factor of 5 between $\alpha$=1 and
5. On the other hand, serious consideration should be given to the
latter possibility of oxygen deficiency and thus metallic
SrTiO$_{3-\delta}$ layers. It is noteworthy that numerous recent
controversial studies and some debate in oxide electronics circles
have centered on the discontinuous IF polarity of LaAlO$_3$/STO
\cite{Ohtomo_LAO_STO} especially regarding the origin of
conduction due to the oxygen vacancies \cite{Eckstein}. As
indicated by the arrow in Fig. 3(a), the range of $n_{sheet}$
values of LaAlO$_3$/SrTiO$_{3-\delta}$ grown at low oxygen
pressure \cite{Kalakukhov} is so wide that it also overlaps our
experimental value of $n_{sheet}\approx3\times10^{14}$ cm$^{-2}$
as well as with the theoretical values \cite{Okamoto1, Kancharla}
of electronic reconstruction at the LTO/STO IF. However, according
to the recent report by Herranz \textit{et al.}, LaAlO$_3$/STO
films were all insulating or highly resistive when cooled in a
high oxygen pressure environment from the deposition $T$ to room
$T$ \cite{Herranz}. We performed a similar \textit{in situ} oxygen
annealing process to compensate for possible oxygen vacancies, but
our LTO/STO
SLs still exhibited metallic behavior. 

\begin{figure}
\includegraphics[width=3.3in,bb=20 20 416 193]{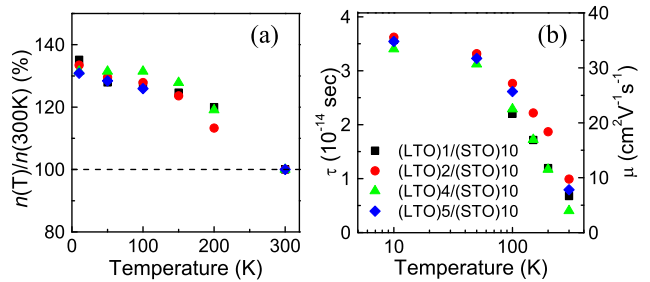}
\caption{\label{fig4} (color online) $T$ dependence of (a) the
carrier density, and (b) the relaxation time and mobility of free
carriers for (LTO)$\alpha$/(STO)10 SLs.}
\end{figure}

Another interesting aspect concerns the strong $T$ dependence of
$n(T)$. Figure 2(c) shows the $T$ dependence of
$\sigma_{1}(\omega)$ for (LTO)2/(STO)10, which are characteristic
of all SLs. Figure 4(a) shows that $n(T)$ increased significantly
by almost 30$\%$ between 300 K and 10 K. The corresponding $n(T)$
in a conventional normal metal is typically of the order of 1$\%$,
even for oxides with strongly correlated electrons. For the
cuprate high-$T_c$ superconductors, for example, it is smaller
than 10$\%$ \cite{Toschi}. Even the opposite $T$ dependence is
observed for doped semiconductors where $n(T)$ decreases as the
carriers become trapped at low $T$. Thus, the increase of $n(T)$
at low $T$ also cannot be explained by the electron-doping effect
of oxygen vacancies of STO. However, we note that the quantum
paraelectric behavior of incipient ferroelectric STO exhibits a
significant increase in dielectric permittivity
($\epsilon_{0}^{STO}$) as $T$ decreases. Therefore, we can explain
the increase of $n(T)$ in the context of electronic reconstruction
at an IF. The charge transfer across the IF of LTO/STO (and thus
$n_{sheet}$) should be affected by the dielectric screening of the
STO layer. The larger value of $\epsilon_{0}^{STO}$ thus gives
rise to the larger screening length of the electronic charges, and
this allows more charges to be transferred across the LTO/STO IF.
(See also the theoretical prediction of the coherent carrier
density as a function of $\epsilon_{0}$ by Kancharla
\cite{Kancharla}.)

Figure 4(b) displays the relaxation time of free carriers
$\tau(T)[=\Gamma_{D}^{-1}(T)]$. As all of our LTO/STO SLs showed a
significantly reduced $\Gamma_{D}$, which is at least one order of
magnitude smaller than a doped Mott system \cite{RCaTiO3}, fairly
high mobility $\mu$(10 K) $\approx$ 35 cm$^2V^{-1}s^{-1}$ was
obtained according to the relation $\mu=e\tau/m^{*}$ with $m^{*}=
1.8m_{e}$. It would be interesting to consider whether signatures
of novel phenomena, such as the quantum Hall effect, may be
observable in the LTO/STO SLs. The measured value of
$\tau\approx3.5\times10^{-14}$ s at 10 K yields
$\omega_{c}\tau=0.01-0.02$, where the cyclotron frequency
$\omega_c=eB/m^{*}$ with the magnetic field $B\approx20$ T and
$m^{*}=1.0-2.5 m_{e}$. From the obtained carrier density and
mobility of our LTO/STO SLs, we determined that the mean free path
of the conducting electrons is about 10 nm ($\approx$ 25 unit
cells long) at 10 K. It is remarkable that there was a recent
observation of the quantum Hall effect in ZnO-based
heterostructures below 1 K \cite{QHE ZnO}, in which the coherence
length of the electron was considerably longer than those in other
oxides.

In summary, our use of IR spectroscopic ellipsometry showed that
all the LTO/STO SLs exhibited a Drude-like metallic response
regardless of the SL periodicity and the individual
layer-thicknesses, even after \textit{in situ} post annealing in
oxygen. Our data yielded a nearly constant $n_{sheet}$ per IF of
$\sim3\times10^{14}$ cm$^{-2}$ with small $\Gamma_{D}$ of
$\sim120$ cm$^{-1}$, and a sizeable mean free path of $\sim$25
unit cells at 10 K. We note that the unusually strong $T$
dependence of $n$ can be interpreted qualitatively as quantum
paraelectric behavior of STO with theoretical prediction of the
leakage of $d$-electrons from LTO to STO across the IF. A picture
of electron-doping by the oxygen vacancy of STO, which has been
the subject of recent debate on similar systems of LaAlO$_3$/STO,
is insufficient to explain our observations in LTO/STO SLs. Hence,
we note the difference between an LTO Mott insulator and an
LaAlO$_3$ band insulator, and suggest that the strongly correlated
$d$-electrons of LTO should play an important role in electronic
reconstruction and the resultant metallic state at a polar (LTO)
and nonpolar (STO) oxide IF.

The authors thank S. Okamoto, A.J. Millis, H.Y. Hwang, G.W.J.
Hassink, J.S. Ahn, D.-W. Kim, Y.S. Lee, J. Yu, K. Char, K.H. Kim
J.H. Park, J.Y. Kim, J.S. Kim, A.V. Boris, and B. Keimer for
valuable discussions, as well as Y.L. Mathis for support at IR-BL
of ANKA. This work was supported by the Creative Research
Initiatives (Functionally Integrated Oxide Heterostructure) of
KOSEF, the Swiss National Science Foundation (SNF project
200021-111690/1), and the Division of Materials Sciences and
Engineering, U.S. Department of Energy (H.N.L.). Experiments in
PLS were supported by MOST and POSTECH.


\begin{thebibliography}{27}
\bibitem{LTO1} For (LTO)1/(STO)10 and (LTO)1/(STO)6, it is rather ambiguous
to define the number of IF. Here, we count two IFs (both sides)
per LTO layer for consistency with other SL samples. Hence, the
$n_{sheet}$ of (LTO)1/(STO)10 and (LTO)1/(STO)6 may be reduced due
to overestimated $N_{I}$.
\bibitem{integral} The values from the theoretical calculations of
$\sim6.5\times10^{14}$ cm$^{-2}$ and $\sim3.4\times10^{14}$
cm$^{-2}$ are obtained by integrating the coherent $d$-electron
density curves displayed in Figs. 2 and 2(a) of Refs.
\cite{Okamoto1} and \cite{Kancharla}, respectively.
\end{thebibliography}
\end{document}